\documentclass[
    ,final            
  ]
  {aipproc}

\layoutstyle{6x9}

\usepackage{amssymb}
\usepackage{fontenc}
\usepackage{times}
\usepackage{mathptmx}
\usepackage{graphicx}

\newcommand{\PWN}{\rm{PWN}}
\newcommand{\PWNe}{\rm{PWNe}}

\newcommand{\hess}{\textit{H.E.S.S.}}
\newcommand{\integral}{\textit{INTEGRAL}}

\newcommand{\xmm}{\textit{XMM-Newton}}
\newcommand{\cxo}{\textit{Chandra-XRO}}
\newcommand{\simbolx}{\textit{Simbol-X}}

\newcommand{\arcmin}{${'}$}
\newcommand{\arcsec}{${''}$}

\newcommand{\aj}{AJ}                   
\newcommand{\apj}{ApJ}

\newcommand{\apss}{Ap\&SS}             
\newcommand{\aap}{A\&A}

\newcommand{\araa}{ARA\&A}
             
\newcommand{\mnras}{MNRAS}             

\begin{document}

\title{The Emerging Population of \\ Pulsar Wind Nebulae in Hard X-rays}



\author{F. Mattana}{
  address={AstroParticule et Cosmologie (APC), CNRS--Universit\'e Paris 7, F-75205 Paris cedex, France}
}

\author{D. G\"otz}{
  address={CEA Saclay, DSM/IRFU/Service d'Astrophysique, F-91191 Gif-sur-Yvette, France}
  ,altaddress={AstroParticule et Cosmologie (APC), CNRS--Universit\'e Paris 7, F-75205 Paris cedex, France} 
}

\author{R. Terrier}{
  address={AstroParticule et Cosmologie (APC), CNRS--Universit\'e Paris 7, F-75205 Paris cedex, France}
}

\author{M. Renaud}{
  address={AstroParticule et Cosmologie (APC), CNRS--Universit\'e Paris 7, F-75205 Paris cedex, France}
}

\author{M. Falanga}{
  address={CEA Saclay, DSM/IRFU/Service d'Astrophysique, F-91191 Gif-sur-Yvette, France}
  ,altaddress={AstroParticule et Cosmologie (APC), CNRS--Universit\'e Paris 7, F-75205 Paris cedex, France} 
}

\begin{abstract}

The hard X-ray synchrotron emission from Pulsar Wind Nebulae probes energetic particles, closely related to the pulsar injection power at the present time. \integral\ has disclosed the yet poorly known population of hard X-ray pulsar/\PWN\ systems. We summarize the properties of the class, with emphasys on the first hard X-ray bow-shock (CTB 80 powered by PSR B1951+32), and highlight some prospects for the study of Pulsar Wind Nebulae with the  \simbolx\ mission.

\end{abstract}

\maketitle


\section{Introduction}

The detection of high-energy radiation from rotation powered pulsars shows that particle acceleration takes place close to the pulsars' magnetospheres  as well as in their surroundings. The particle stream initiated by the electrostatic magnetospheric gaps constitutes a magnetized wind, which flows relativistically until it experiences the confinement by the surrounding medium, be it the remnant of the supernova explosion or the interstellar gas. In this interaction a strong shock is produced, which terminates the wind bulk motion and re-accelerates the electrons. The presence of the pulsar wind is so revealed in a Pulsar Wind Nebula (\PWN), which shines along all the electromagnetic spectrum: the synchrotron process yields photons from the radio band up to the energies of several MeV, while the inverse Compton scattering of the ambient radiation field yields higher energy photons, up to tens of TeV \citep{Gaenslerslane06}. 

As the nebular radiation arises from the integrated history of particle injection by the pulsar convolved with the interaction with the environment, \PWNe\ store, and slowly release, most of the pulsar rotational kinetic power, the so-called spin-down luminosity $\dot{E}$. However, the synchrotron and inverse Compton radiations can trace particles of different energies and ages \cite{Mattana09a}, hence the combination of X- and $\gamma$-ray measurements may allow to disentangle the contribution of the pulsar from the one of its environment. The high energy tail of the synchrotron emission has a special role in this respect, as it is the most closely related to the pulsar: it probes the most energetic particles, up to PeV energies, with short lifetime (for instance, the cooling time for electrons radiating 50 keV photons is $\sim$ 50 yr -- 1 kyr). At such energies, deviations from pure power law spectra might be induced by the evolution of the spin-down luminosity, by the radiative cooling, or by the attainment of the maximum acceleration energy.

Although $\sim$50 \PWNe\  are observed in radio waves, soft X-rays (around 0.5--10 keV), and very high energy (VHE, $E > 0.1$ TeV) $\gamma$-rays, only a few are detected in the hard X-ray band (E $\gtrsim$ 20 keV). The sample of hard X-ray  \PWNe\ is increasing thanks to the unprecedented sensitivity and imaging capability (6\arcmin~ angular resolution, HWHM) of the  IBIS/ISGRI camera \cite{Lebrun03} on board the 
\integral\ satellite, which has performed a survey of the Galactic disk  from 20 keV up to several
hundreds of keV. Taking advantage of the long exposure time accumulated since the launch of the mission
(up to 16 Msec in the Galactic Center),  the IBIS/ISGRI survey has revealed a dozen of young pulsar/PWN complexes among a few hundreds of compact binary systems and unidentified objects. Here we summarize their properties, and outline the opportunities for the study of \PWNe\ with the \simbolx\ observatory.

\section{Pulsar/PWN systems detected by IBIS/ISGRI}

Table \ref{tab:1} lists the pulsars/\PWNe\  detected by IBIS/ISGRI. These sources are powered by very young (Crab-like) or youngish (Vela-like) pulsars. For most of them, the hard X-ray flux is dominated by the nebular emission, with the exceptions of PSR B1509-58, PSR J1617-5055, and PSR J1838-0655, whose flux is dominated by the pulsed magnetospheric emission. Two hard and stable unidentified \integral\ sources also proved to be linked to young pulsars and \PWNe: IGR J18135-1751 \cite{Ubertini05} and IGR J18490-0000 \cite{Terrier08}. Their identification is supported by the detection of a TeV counterpart, like the large majority of identified  pulsar/\PWN\ systems. 

\begin{table}[!h]
\centering
{\footnotesize
\vspace*{0.5cm}
\begin{tabular}{llcrcll}
\hline \hline
\noalign{\smallskip}
PSR & PWN       & $\dot{E}$        & $\tau_c$ & Dominant  & \hess\       & Refs.\\
        &                & (erg s$^{-1}$) & (kyr)         & Component       & source      &          \\[3pt]
\hline
\noalign{\smallskip}
B0531+21     &  Crab                    & $4.7\times 10^{38}$      & 1.24           & PWN       &  HESS J0534+220      & \cite{Mineo06}\\ 
B0540-69      &  N158A                 & $1.5\times 10^{38}$      & 1.67           & PWN       &  -                                & \cite{Campana08}\\ 
B0833-45      &  Vela                    & $6.9\times 10^{36}$      & 11.3            & PWN       &  HESS J0835-455      & \cite{Mangano05}\\ 
J1420-6048  & K3 in Kookaburra & $1.0\times 10^{37}$      & 13.0         & ?    & HESS J1420-607   & \cite{Hoffmann07} \\
B1509-58      &  MSH 15-52         & $1.8\times 10^{37}$      & 1.55           & PSR        &  HESS J1514-591      & \cite{Forot06}\\ 
J1617-5055   &  underluminous          & $1.6\times 10^{37}$      & 8.15           & PSR &  HESS J1616-508 ?   & \cite{Landi07}\\ 
J1811-1925   &  G11.2-0.4           & $6.4\times 10^{36}$      & 23.3         & PWN        &  -                               & \cite{Dean08a}\\  
J1833-1034   &  G21.5-0.9           & $3.4\times 10^{37}$      & 4.85            & PWN       &  HESS J1833-105     & \cite{Derosa09}\\ 
J1838-0655   & AX J1838.0-0655 & $5.5\times 10^{36}$      & 22.7          & PSR       &  HESS J1837-069     & \cite{Gotthelf08}\\
J1846-0258   &  Kes 75                & $8.3\times 10^{36}$      & 0.73              & PWN       &  HESS J1846-029     &\cite{Mcbride08}\\ 
B1951+32     &  CTB 80                & $3.7\times 10^{36}$      &  107.      & PWN & -     & \cite{Mattana09b} \\
\noalign{\smallskip}
\noalign{\smallskip}
-                    & IGR J18135-1751    & - & - & PWN & HESS J1813-178 & \cite{Ubertini05} \\
-                    & IGR J18490-0000    & - & - & PSR? & HESS J1849-000 & \cite{Terrier08} \\
\noalign{\smallskip}
\hline \hline
\end{tabular}
\caption{\small Hard X-ray pulsar/PWN systems detected by IBIS/ISGRI. 
\label{tab:1}}
}
\end{table}

\begin{figure}
\begin{tabular}{cc}
 \includegraphics[width=.59\textwidth, height=6.45cm]{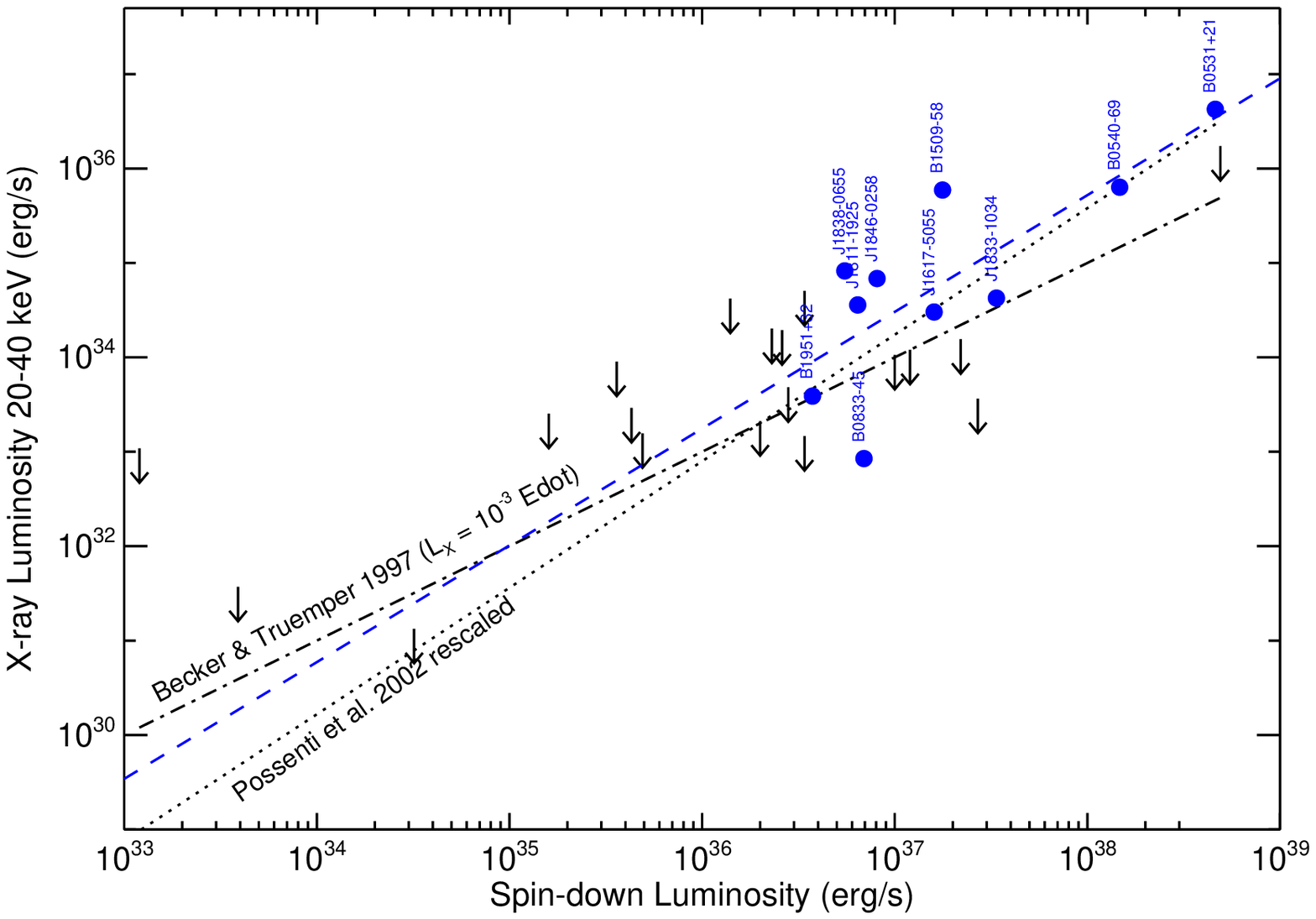} &
 \includegraphics[width=.39\textwidth, height=6.1cm]{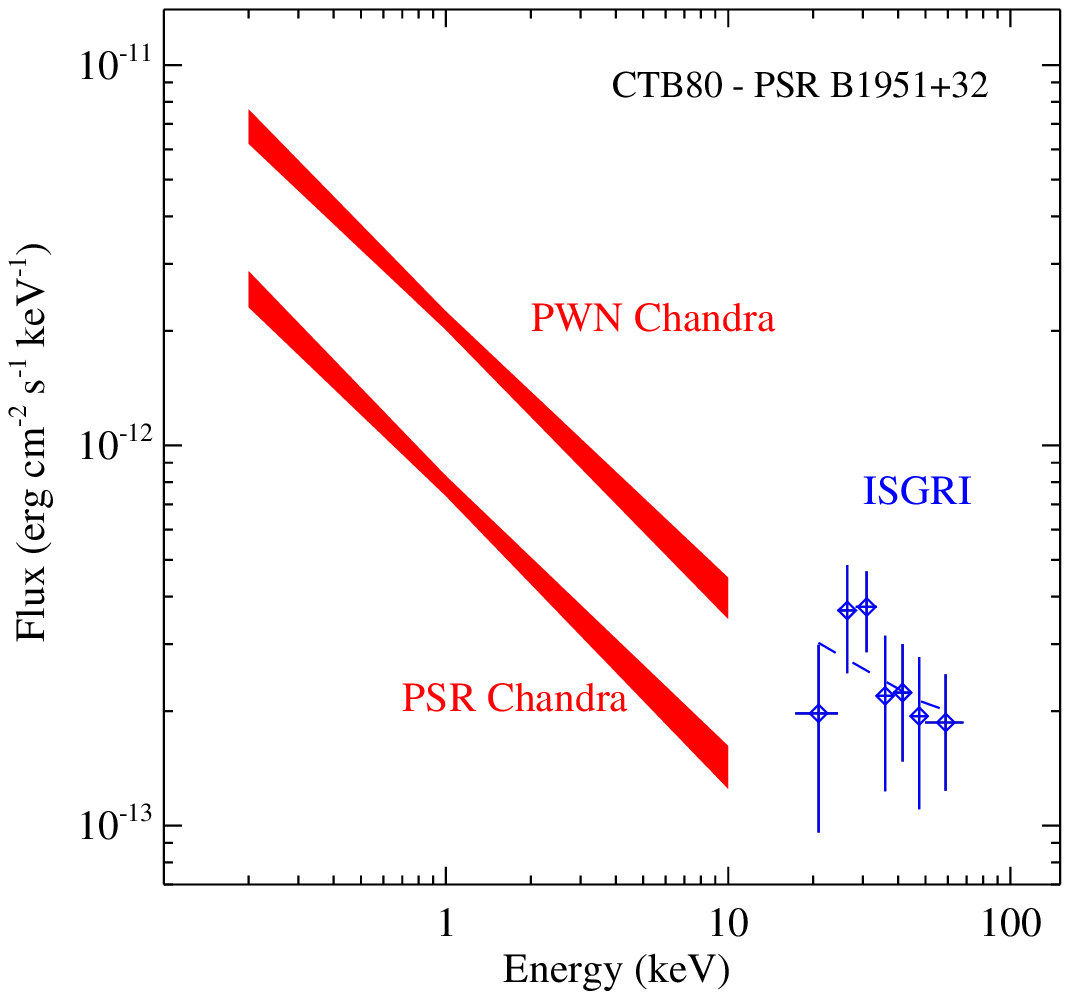} 
\end{tabular}
\caption{
Left panel: hard X-ray luminosity (20--40 keV) versus spin-down luminosity $\dot{E}$ 
for the pulsar/PWNe systems detected by IBIS/ISGRI (Table \ref{tab:1}). 
Dashed line: unweighted least square fit.Two relations for the soft X-ray band are 
shown for comparison: the relation by \cite{Possenti02} rescaled in the 20--40 energy 
band (dotted line), and the simple scaling $L_X = 10^{-3} \dot{E}$ by \cite{Becker97} 
(dot-dashed line). 
Right panel: IBIS/ISGRI spectrum of CTB 80/PSR B1951+32 \cite{Mattana09b}. The source is significant up to 70 keV. The \cxo\ spectra \cite{Moon04} of the pulsar and of the \PWN\ are also shown.
\label{fig1}}
\end{figure}

The luminosity in the 20--40 keV energy band of the sources in Table \ref{tab:1} is found to increase with the spin-down luminosities $\dot{E}$ (Figure \ref{fig1}, left panel). This $L_X-\dot{E}$ scaling relation is similar to the one known in the soft X-ray band for pulsars as well as for  \PWNe\ \cite[e.g.,][]{Possenti02, Kargaltsev08}. The fact that also the hard X-ray luminosities scale with $\dot{E}$ is not surprising, provided that the spectra in the 20--40 keV energy band smoothly extrapolate the soft X-ray spectra. We also considered the 3$\sigma$ flux upper-limit at the position of other 19 pulsars powering a soft X-ray \PWN\ (assuming a Crab-like spectrum). Some of the upper-limits are well below the unweighted least square fit relation ($\log_{10} L_{\footnotesize{\textrm{20-40 keV}}} = 34.5 + 1.24 \, \log_{10} \dot{E}_{37}$, where $\dot{E} = \dot{E}_{37} \times 10^{37}$ erg s$^{-1}$), but still consistent with the scatter between the measured fluxes and the best fit relation, which can be larger than one order of magnitude. For all the detected pulsars/\PWNe\ $\dot{E} > 10^{36}$ erg s$^{-1}$; this apparent spin-down threshold  can be due to lack of sensitivity.

The hint that also more evolved pulsar/\PWN\ systems may be detected is supported by the recent discovery of CTB 80 powered by the middle-aged PSR B1951+32. While bright in $\gamma$-rays above 100 MeV \cite{Ramanamurthy95}, PSR B1951+32 is X-ray faint, yielding a detection of its pulsation at a 99\%  significance level \cite{Safiharb95}. The bulk of the X-ray flux originates from the \PWN\ confined by the supersonic motion of the pulsar in the associated supernova remnant CTB80. After $\sim$50 kyr since its birth \cite[dynamical age,][]{Zeiger08}, the pulsar has reached the boundaries of the supernova ejecta, yielding a bow-shock nebula in X-rays, H$\alpha$ \cite{Moon04}, and radio band \cite{Castelletti03}. The comparison of the IBIS/ISGRI and \cxo\ spectra (Fig. \ref{fig1}, right panel) shows that the \PWN\ provides the main contribution also to the hard X-ray flux. Therefore CTB 80 is the first bow-shock detected in hard X-rays.

\section{Prospects for \textit{Simbol-X}}
 \integral\ has disclosed the brightest members of the population of hard X-ray pulsar/\PWN\ systems. However, the flux above 20 keV of the lately discovered sources is close to the IBIS/ISGRI sensitivity limit (a few mCrab), and new detections are currently hampered by the systematics in the image reconstruction. Most of all, a better angular resolution is needed to disentangle the pulsar from the \PWN\ contributions. The focussing technique of \simbolx\ aims at overcoming these limitations.

We can roughly estimate the number of pulsar/\PWNe\ systems that \simbolx\ might detect. Using the current
\simbolx\ simulation tools, the errors on the measured spectral parameter of a source with flux 
$F_{\textrm{\footnotesize 20-70 keV}} = 5 \times 10^{-13}$ erg cm$^{-2}$ s$^{-1}$ ($\sim$ 10~$\mu$Crab) would be of the order of 20\% (assuming an exposure time of 100 ks, and $N_H =3 \times 10^{22}$ cm$^{-2}$). Hence, using the $L_X -\dot{E}$ relation as a luminosity predictor and accounting for the measured distances, $\sim$80 pulsar/\PWN\ systems are expected to have an higher flux.

However, \simbolx\ will perform selected pointings instead of a survey. By taking advantage of a PSF $\sim$ 15\arcsec\ in diameter, comparable to the one of \xmm, \simbolx\ will be able to resolve the most extended \PWNe, and to access the dynamics of the nebula at higher energies. Recently \cite{Uchiyama08} using \textit{Suzaku}/XIS found that G18.0-0.1, powered by PSR B1823-13,  is extended up to 15\arcmin\ in 1--3 keV. Since a significant early expansion combined with a very low magnetic field implies a bright hard X-ray nebula, the combination of spatially resolved X-ray and $\gamma$-ray measurements would allow to constrain the nebular magnetic field and the maximum particle energy.






\bibliographystyle{aipproc}   




\end{document}